\UseRawInputEncoding 
\documentclass[aps,pra,twocolumn,showpacs,showkeys,superscriptaddress,reprint]{revtex4-2} 
\usepackage{graphics} 
\usepackage{graphicx} 
\usepackage{amsmath} 
\usepackage{color}

\begin{document} 

\title{Spin-selective insulators} 
\author{J. Silva-Valencia} 
\email{jsilvav@unal.edu.co} 
\affiliation{Departamento de F\'{\i}sica, Universidad Nacional de Colombia, A. A. 5997 Bogot\'a, Colombia.} 

\date{\today} 

\begin{abstract}
Spin-selective insulators emerge in systems composed of fermions with two internal degrees of freedom and another carrier, which could be fermionic or bosonic. These insulators are characterized by a gapless state for one kind of fermion and an insulator state for the other, with the latter satisfying a commensurability relation that involves the other carrier. We review the different scenarios where these unique insulators arise, focusing on Bose-Fermi mixtures, the most recent and promising scenario for observing these insulators in cold atom setups. 
\end{abstract} 


\maketitle 
\tableofcontents

\section{\label{sec1}Introduction}
The main reason that condensed matter physics exhibits such plethora of physical phenomena is that it involves a great number of carriers that undergo strong interactions in various degrees of freedom, such as charge, spin, lattice, orbital, topology, and disorder, which generate diverse ground states and intriguing quantum phenomena. A basic description of materials in terms of weakly interacting electron systems and the use of the filling of the electronic bands allowed us to distinguish between metals and insulators; however, experiments suggested the need to include an on-site Coulomb interaction between electrons in order to explain some results and the \textit{metal-insulator transition}~\cite{Mott-RMP68}. The interplay between the kinetic energy and the local repulsive interaction leads to a half-filling insulator with a charge gap, where at each site there is one electron forming a global antiferromagnetic state called \textit{Mott-insulator}~\cite{Imada-RMP98,Guan-RMP13}. Note that the bosonic counterpart also exhibits insulator states at integer densities, which has been experimentally observed~\cite{Cazalilla-RMP11}.\par 
The diverse degrees of freedom of the carriers enable the emergence of other insulators, such as charge density waves~\cite{Gruner-RMP88} and spin density waves~\cite{Gruner-RMP94}, among others. Taking into account the topology of the band structures allows us to classify insulators into different classes, and under the influence of strong spin-orbital coupling, the usual ordering conduction and valence bands of an ordinary insulator can be inverted, allowing the appearance of metallic surface states. This illustration indicates that there are gapless surface states inside the bulk energy gap, which characterizes the topological insulators, where these surface states exhibit a Dirac cone–type dispersion~\cite{Hasan-RMP10,XLQi-RMP11,BYan-AR17}.\par
In iron chalcogenide materials, the physical properties depend on the competition between the electron correlation strength and the nature of the Fermi surface, which causes a subset of orbitals (denoted “heavy”) to have a much larger effective mass than another group (denoted “light”), allowing the possibility that the heavy electrons are Mott localized and coexist with the itinerant light electrons, a unique insulator state called an orbital-selective Mott phase~\cite{Anisimov-EPJB02,MYi-NC15}.\par 
As mentioned before, there are many intriguing insulators with interesting properties; however, in this review, we consider the spin-selective insulators, which have emerged so far in two different scenarios: the Kondo lattice model and Bose-Fermi mixtures. The common element between the two scenarios is a system of fermions with two internal degrees of freedom that coexist with other systems of localized spins or bosons. The strong interactions generate a state characterized by a gapless background due to one kind of fermions, while the other kind of fermions remains in an insulator state, fulfilling a commensurate relation with the other ingredient (fermions or bosons), which is the reason it is called a spin-selective insulator. Although  both scenarios will be discussed, special attention will be dedicated to mixtures, which are projected to be  the most promising platform for observing these states.\par 
The plan of this article is as follows: The original scenario where the spin-selective insulator emerges, which is the Kondo lattice model, is discussed in Sec.~\ref{sec2}. The recent Bose-Fermi mixture scenario will be covered extensively in Sec.~\ref{sec3}. In Sec.~\ref{sec4}, we will suggest possible new scenarios for the emergence of spin-selective insulators. Finally, some conclusions will be put forth in Sec.~\ref{sec5}
\section{\label{sec2}Two-band Fermionic models}
Strong correlations between electrons generate interesting and intriguing experimental results in diverse kind of materials. For instance, in intermetallic compounds containing rare-earth or actinide elements, the strong electron-electron correlation is crucial. In these materials, the linear coefficient of the specific heat and the Pauli spin susceptibility are extremely large compared to that of conventional metals, keeping the Wilson ratio around the unity and enabling a description in terms of the Fermi-liquid theory. To explain the experiments, it is common to consider a quasiparticle mass around two to three orders of magnitude larger than the bare electron mass, giving the name of \textit{heavy fermions} to these materials~\cite{Hewson-Book}.\par 
\begin{figure}[t]
{\centering \resizebox*{2.9in}{!}{\includegraphics{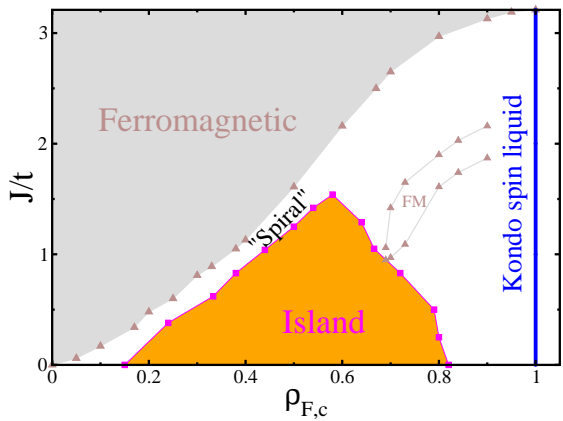}} \par}
\caption{\label{fig1} Zero-temperature phase diagram of the one-dimensional Kondo lattice model, interaction coupling ($J/t$) versus conduction electron density ($\rho_{F,c}$). This summarizes the results known to date for the borders between phases~\cite{Tsunetsugu-RMP97,Shibata-JPCM99,Gulacsi-AP04,Caprara-EPL97,Honner-PRL97,McCulloch-PRB02,Basylko-PRB08}}
\end{figure}
Two different kind of electrons determine the physics of the heavy-fermion materials. One is a set of electrons in the inner $f$ orbitals, which remains localized even in a periodic lattice. The other set consists of the conduction electrons in $s$-, $p$-, or $d$-atomic orbitals, which moves through a lattice. The local interaction between $f$ and conduction electrons generates the largest energy scale of the system, followed by the Hund’s-rule coupling~\cite{Tsunetsugu-RMP97}.\par 
The heavy-fermion materials are modeled as strongly coupled conduction electrons and nearly independent localized spins, which generate a variety of ground states, such as the normal heavy-fermion state, unusual superconducting states, antiferromagnetically ordered states, and topological Kondo insulators~\cite{Gegenwar-NP08}.\par 
Although the study of the physics of block heavy-fermion materials is still current and offering surprises, a new impetus has emerged with the possibility of creating heterostructures based on heavy-fermion materials, for instance superlattices with a unit cell composed of a two-dimensional heavy-fermion material and a different material exhibit intriguing superconducting and magnetic properties, which are tunable by changing the superlattice structure, opening the possibility of creating new functional devices based on heavy-fermions~\cite{Shishido-S10,Mizukami-NP11,Shimozawa-RPP16}.\par 
Another branch, opened in the last decade, relate to heavy fermions is the possibility of emulating their main ingredients in clean and fully controllable setups, which is achieved by confining alkaline-earth-like atoms (Yb and Sr) in optical lattices~\cite{Gorshkov-NP10,Riegger-PRL18,KOno-PRA19}.\par 
Recently, it was observed that the material CeCo$_2$Ga$_8$ exhibits a strong anisotropy ratio of its magnetic exchange interactions, which allows concluding that this is a realistic example of a Kondo chain, thus harmonizing the theoretical results and the experimental ones~\cite{KCheng-PRM19}.\par 
\begin{figure}[t]
{\centering \resizebox*{3.0in}{!}{\includegraphics{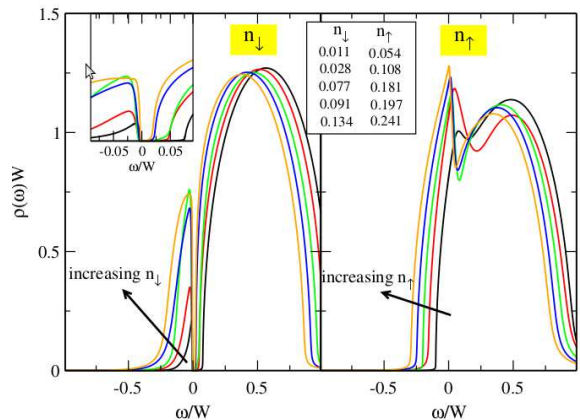}} \par}
\caption{\label{fig2} Local spin-resolved spectral functions for the Kondo lattice model in a Bethe lattice. The interaction coupling is $J/4t=0.25$, hence the ground state is ferromagnetic. In the inset, we can see the gap for the spin-down component. Reproduced from Ref.~\cite{Peters-PRL12}.}
\end{figure}
As mentioned before, the spin exchange between conduction electrons and localized impurities is relevant for understanding heavy-fermions, which can lead us to an indirect exchange interaction between impurities called Rudermann–Kittel–Kasuya–Yosida (RKKY) interaction, which  emerges due to the fact that the impurities are coupled to the same conduction electrons, making possible a magnetic interaction between impurities and a magnetic ordering throughout the system. A different scenario involves a collective screening of the impurities by the conduction electrons called the Kondo effect. The interplay between the RKKY interaction and the Kondo effect is crucial here~\cite{Hewson-Book}. The main models considered to explain the experimental results of heavy-fermion materials are the periodic Anderson model and the Kondo lattice model, the latter  being a simpler model obtained from the former one. The Kondo lattice model considers a kinetic energy term and a local spin-spin interaction term between the localized impurity and the conduction electrons, and its Hamiltonian reads
\begin{equation}
\label{HKLM}
 \hat{H} = J\sum _{i}\mathbf{\hat{S}}_{i}\cdot\mathbf{\hat{s}}_{i}-t\sum _{<i,j>,\sigma}\left( \hat{c}^{\dagger }_{i,\sigma }\hat{c}_{j,\sigma }+ \text{H.c.}\right),
\end{equation}
\noindent where \( \hat{c}^{\dagger }_{i\sigma }\) creates a conduction electron at site \( i \) with spin \(\sigma \) so that $\hat{n}^{Fc}_{i,\sigma}=\hat{c}_{i,\sigma}^{\dag}\hat{c}_{i,\sigma}$ is the local number operator. $\mathbf{\hat{S}}_{i}$ represents a localized spin-1/2 operator, while $\mathbf{\hat{s}}_{i}=\frac{1}{2}\sum_{\alpha\beta}\hat{c}^{\dagger }_{i,\alpha }\boldsymbol{\sigma}_{\alpha\beta}\hat{c}_{i,\beta }$ is the spin operator of a conduction electron, $\boldsymbol{\sigma}$ being the vector of the Pauli matrices. $J$ is the antiferromagnetic local exchange coupling, and $t$ is the hopping integral between nearest-neighbor sites.\par 
Despite its apparent simplicity and the arduous study on the Kondo lattice model, which has involved multiple techniques, it constantly gives us new surprises that keep it alive and current even today~\cite{Tsunetsugu-RMP97,Shibata-JPCM99,Gulacsi-AP04,Caprara-EPL97,Honner-PRL97,McCulloch-PRB02,Basylko-PRB08}. In Fig. \ref{fig1}, we sketch the one-dimensional phase diagram of the Kondo lattice model, the local exchange coupling ($J/t$) versus the density of the conduction electrons $\rho_{F,c}=N_F/L$, where $N_F$ is the total number of conduction electrons and $L$ is the lattice size. When the number of conduction electrons matches the lattice size, the half-filling configuration ($\rho_{F,c}$=1) is reached and a local singlet is expected at each site. This ground state is an insulator with nonzero spin and charge gaps, which is called a Kondo spin liquid. A richer scenario emerges far away from half-filling with a main ferromagnetic phase dominating the diagram, although it decreases when the density increases. A phase with zero spin gap and a quasi-long-range order that depends on the density is established for lower values of the local exchange, which is called the ``island" phase~\cite{Garcia-PRL00,Garcia-PRB02,Garcia-PRL04,Xavier-PRL03}. Between the island and ferromagnetic phases, a ''spiral`` region with unclear magnetic order in the system emerges~\cite{Garcia-PRL04}. An additional small ferromagnetic phase with intermediate coupling strengths and densities $\rho_{F,c} > 2/3$ has been reported~\cite{Peters-PRB12,McCulloch-PRB02}.\par
\begin{figure}[t]
{\centering \resizebox*{3.4in}{!}{\includegraphics{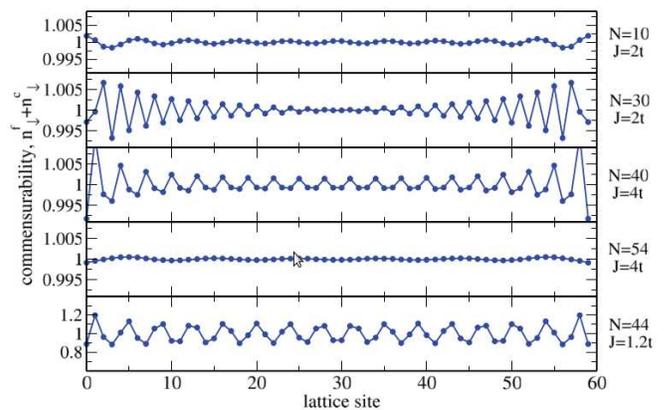}} \par}
\caption{\label{fig3} Local commensurability profile along the lattice for a 1D Kondo lattice model. A lattice with 60 sites was considered and each panel shows the evolution of the local commensurability for a specific ferromagnetic state characterized by the parameters on the right. Lines are a guide for the eye. Reproduced from Ref.~\cite{Peters-PRB12}.}
\end{figure}
The ferromagnetic phase contains interesting physics, as was shown by Peters \textit{et al.} ~\cite{Peters-PRL12}, who used dynamical mean-field theory to solve the Kondo lattice model, considering a two-dimensional square lattice and a Bethe lattice. The local spin-resolved spectral functions reported for the latter lattice appear in Fig. \ref{fig2}, where they considered an antiferromagnetic Kondo coupling $J/W=0.25$ (bandwidth $W=4t$), for which the ferromagnetic phase extends to the conduction density $\rho_{F,c}=\rho^{\uparrow}_{F,c}+\rho^{\downarrow}_{F,c}=0.5$. The surprising result is that the spectral function for the majority-spin ($\rho^{\uparrow}_{F,c}$) is different from that for the minority-spin ($\rho^{\downarrow}_{F,c}$). The minority-spin spectral function exhibits a gap at the Fermi energy, while that for the majority-spin has a peak at the Fermi energy $\omega=0$ and a dip for $\omega>0$, which suggests that despite the ferromagnetic state being metallic, one kind of conduction electron contributes little to the low-temperature behavior and the responsibility is assumed by the other kind. A further increase in the occupation number leads to the dip becoming larger and closer to the Fermi energy, which will take us to a paramagnetic state.\par 
An insulator state called spin-selective Kondo insulator is established for the minority-spin conduction electrons, which has not yet been explained. To elucidate this, Peters \textit{et al.} considered that the localized spins are formed by a half-filled $f$ orbital with a strongly local repulsion interaction between the electrons, so that the spin polarization is given by $\langle S_{z,i} \rangle=0.5\bigl(\langle n^{Ff}_{i,\uparrow}\rangle + \langle n^{Ff}_{i,\downarrow}\rangle\bigr)$. In Fig. \ref{fig3}, we display the local commensurability profile along a one-dimensional lattice, calculated using the density matrix renormalization group (DMRG) method with open boundary conditions~\cite{Peters-PRB12}. The local commensurability is given by $\langle n^{Fc}_{i,\downarrow}+n^{Ff}_{i,\downarrow}\rangle =  \langle n^{Fc}_{i,\uparrow}\rangle -\langle S_{z,i} \rangle +0.5$, and its evolution is shown for a lattice with 60 sites and diverse numbers of conduction electrons and antiferromagnetic couplings. In all the panels, we observe that the local commensurability oscillates around $\langle n^{Fc}_{i,\downarrow}+n^{Ff}_{i,\downarrow}\rangle \sim 1$, but the total number of minority-spin conduction electrons plus spin-down $f$ electrons is commensurate with the lattice size. The oscillations are due to the open boundary conditions, and despite them, the local commensurability condition is fulfilled within 1$\%$ for the upper four panels, which correspond to states inside the main ferromagnetic phase, while the lower panel, related to the smaller ferromagnetic area in the phase diagram, exhibits very strong oscillations. It is important to mention that the above commensurability was also observed for the Kondo lattice model in a Bethe lattice and a two-dimensional square lattice; therefore the conclusion is that within the ferromagnetic state the following commensurability condition is fulfilled:
\begin{equation}
\label{SScondition}
 \rho^{\downarrow}_{F,c}+\rho^{\downarrow}_{F,f}=1.
\end{equation}
Taking into account that the half-filling condition for the localized band requires that $\rho^{\uparrow}_{F,f}+\rho^{\downarrow}_{F,f}=1$, it is clear that $\rho^{\uparrow}_{F,f}=\rho^{\downarrow}_{F,c}$. Another important point is that the majority spins do not fulfill any commensurability relation. Note that the minority-spin conduction electrons and a part of the $f$ electrons collaborate to form a Kondo singlet, while the rest give the ferromagnetic character to the state.\par 
To summarize, it was found that in the ferromagnetic metallic phase of the Kondo lattice model with antiferromagnetic coupling, the majority-spin conduction electrons are in a gapless state, while the minority-spin ones remain in an insulator state, which has an associated a commensurability relation. This insulator was referred to as a spin-selective Kondo insulator, for us simply a spin-selective insulator, and this was the first scenario in which it arose.\par 
We would like to point out that Bazzanella and Nilsson also explain the ferromagnetic phase and the emergence of the spin-selective insulators through a canonical transformation that expresses the Kondo lattice model in terms of Majorana fermions~\cite{Bazzanella-PRB14}.\par 
From the above discussion, we know that the Kondo lattice model can offer still new physics and phenomena, which is evident when we consider the new possibilities that emerge from the attempts to emulate the Kondo lattice model in diverse environments~\cite{JSV-PRA12,JSV-EPJB12a,RCCaro-JMMM20,JSV-PRB01,JSV-JPCM01}.\par
A spin-selective insulator also emerges in the spin-asymmetric Hubbard model in a partially filled Lieb lattice with spin-dependent electron band dispersions, which could be relevant to the electronic states at the LaAlO$_3$/SrTiO$_3$ interface~\cite{Faundez-PRB18}.\par 

\section{\label{sec3}Bose-Fermi mixtures}
Studying a system composed of a mixture of carriers that obey the Bose-Einstein and Fermi-Dirac statistics was a utopia until the emergence and evolution of the ultracold atom field, which has allowed verifying and extending several ideas and concepts in physics in clean and fully controllable setups~\cite{IBloch-RMP08,Esslinger-AR10,IBloch-NP12,Gross-S17}. Mixing fermionic and bosonic isotopes of the same or different atoms, experimentalists have created unimaginable mixtures, controlling the number of each kind of carrier, the interspecies and intraspecies interactions~\cite{Truscott-S01,Schreck-PRL01,Hadzibabic-PRL02,Roati-PRL02,Ott-PRL04,Silber-PRL05,Gunter-PRL06,Ospelkaus-PRL06,Zaccanti-PRA06,McNamara-PRL06,Best-PRL09,Fukuhara-PRA09b,Deh-PRA10,Tey-PRA10,Sugawa-NP11,Schuster-PRA12,Tung-PRA13,Ferrier-Barbut-S14,Delehaye-PRL15,Vaidya-PRA15,XCYao-PRL16,Onofrio-PUsp16,YPWu-JPB17,Roy-PRL17,Schafer-PRA18}. As expected, new and exciting phenomena have emerged, such as a Bose-Fermi superfluid mixture~\cite{Trautmann-PRL18}, phase separation~\cite{Lous-PRL18}, and attractive interaction between bosons mediated by fermions~\cite{DeSalvo-Nat19}.\par 
The description of mixtures of bosons and fermions, taking into account the interactions, leads us to the Bose-Fermi-Hubbard model, for which there is some level of approximation, and it has been widely studied using analytical and/or numerical techniques~\cite{Albus-PRA03,Cazalilla-PRL03,Lewenstein-PRL04,Mathey-PRL04,Roth-PRA04,Frahm-PRA05,Batchelor-PRA05,Takeuchi-PRA05,Pollet-PRL06,Mathey-PRA07,Sengupta-PRA07,Mering-PRA08,Suzuki-PRA08,Luhmann-PRL08,Rizzi-PRA08,Orth-PRA09,XYin-PRA09,Sinha-PRB09,Orignac-PRA10,Polak-PRA10,Mering-PRA10,Anders-PRL12,Masaki-JPSJ13,Bukov-PRB14,TOzawa-PRA14,Bilitewski-PRB15}. Among the diverse ground states found, the mixed Mott insulator stands out and is characterized by the commensurability between the lattice size and the total number of carriers (bosons + fermions), which was predicted for a mixture of polarized bosons and fermions~\cite{Zujev-PRA08} and experimentally observed in a mixture of ytterbium atoms confined in a three-dimensional optical lattice~\cite{Sugawa-NP11}.\par 
An intermediate approach for describing a Bose-Fermi mixture consists of taking into account the internal degrees of freedom of fermions and treating the bosons as scalars, whose Hamiltonian is given by
\begin{equation}\label{eq:HBF}
	\hat{H}_{BF}=\hat{H}_F+\hat{H}_B+\hat{H}_I,
\end{equation}
\noindent where $\hat{H}_B$, $\hat{H}_F$, and $\hat{H}_I$ represent the contribution of bosons, fermions, and the interaction between fermions and bosons, respectively.\par 
$\hat{H}_F$ corresponds to the Fermi-Hubbard Hamiltonian, which is given by 
\begin{equation}\label{eq:HF}
\hat{H}_F = -t_F \sum_{\langle i,j\rangle,\sigma}\left(\hat{f}_{i\sigma}^\dagger\hat{f}_{j\sigma} + \text{H.c.} \right) + \frac{U_{FF}}{2}\sum_{i,\sigma\neq\sigma'}\hat{n}_{i,\sigma}^F\hat{n}_{i,\sigma'}^F.
\end{equation}
Here, $\hat{f}_{i,\sigma}$ $(\hat{f}_{i,\sigma}^\dagger)$ annihilates (creates) a fermion with spin $\sigma=\uparrow,\downarrow$ at the lattice site $i$, and $\hat{n}^{F}_{i,\sigma}=\hat{f}_{i,\sigma}^{\dag}\hat{f}_{i,\sigma}$ is the local number operator for each kind of fermions, such that $\hat{n}^{F}_{i}=\hat{n}^{F}_{i,\uparrow}+\hat{n}^{F}_{i,\downarrow}$. The hopping parameter between nearest-neighbor sites ($\langle i,j \rangle$) for fermions is $t_F$, and $U_{FF}$ quantifies the local fermion-fermion interaction. The global density for $\sigma$-fermions is defined as $\rho^{\sigma}_F=N^{\sigma}_F/L$, where $N^{\sigma}_F$ is the number of fermions with spin $\sigma$. $\rho_F=\rho^{\uparrow}_F+\rho^{\downarrow}_F$ is the total fermionic density, which varies from zero to two, $\rho_F=1$ being the half-filling configuration.\par 
\begin{figure}[t] 
\includegraphics[width=18pc]{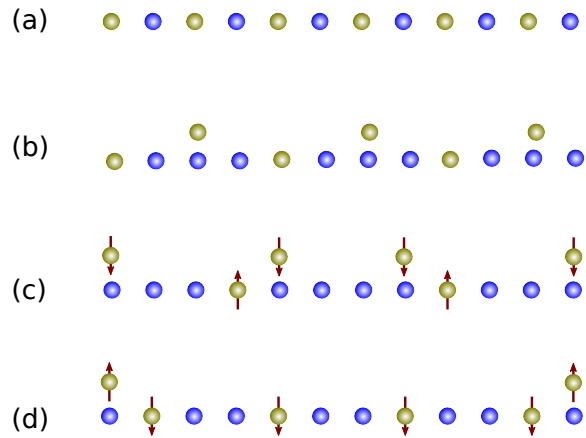} 
\caption{\label{fig4} Sketches of possible distributions of fermions (golden) and bosons (blue) in a 12-sites lattice. (a) Mixed Mott insulator state, for which $\rho_F+\rho_B=1$. (b) Mixed noncommensurate insulator for a balance mixture ($\rho_B+\tfrac{1}{2}\rho_F=1$). (c) and (d) show insulators for an imbalanced mixture with $I=1/3$.}
\end{figure} 
The Bose-Hubbard Hamiltonian $\hat{H}_B$ corresponds to
\begin{equation}\label{eq:HB}
	\hat{H}_B = -t_B \sum_{\langle i,j\rangle}\left(\hat{b}_i^\dagger\hat{b}_j + \text{H.c.} \right)+\frac{U_{BB}}{2}\sum_{i}\hat{n}_{i}^{B}\left(\hat{n}_{i}^{B}-1\right),
\end{equation}
\noindent where the operator $\hat{b}_i^\dagger$ ($\hat{b}_i$) creates (annihilates) a boson at the lattice site $i$, and  $\hat{n}_i^B=\hat{b}_i^\dagger\hat{b}_i$ is the local number operator. The interaction between bosons is quantified for the parameter $U_{BB}$, while the hopping of the bosons is modulated by $t_B$. $N_B$ is the number of bosonic atoms, and $\rho_B=N_B/L$ is the global density of bosons, which varies from zero to one.\par 
\begin{figure}[t] 
\includegraphics[width=18pc]{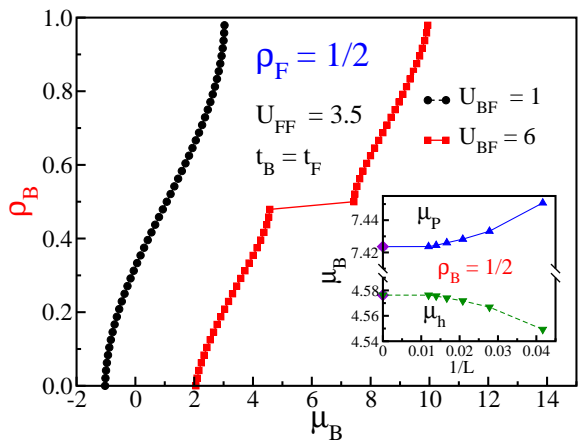} 
\caption{\label{fig5} Evolution of the chemical potential $\mu_B$ with the bosonic density $\rho_B$, for a mixture with  $\rho_F = 1$ (half-filling) and $U_{FF}=3.5$. Two different values of the boson-fermion interaction $U_{BF}=6$ (red) and $U_{BF}=1$ (black)
were considered. Inset: Chemical potential as a function of the inverse of the lattice size at the bosonic density $\rho_B = 1/2$. The diamonds determine a finite gap of $\Delta^B=2.85$. The lines are visual guides. From ~\cite{Avella-PRA19}.} 
\end{figure} 
Considering the above definitions, the interaction between two-color fermions and scalar bosons is described by 
\begin{equation}\label{H-inter}
	\hat{H}_{I}=U_{BF}\sum^{L}_{i=1}\hat{n}_i^B\left(\hat{n}_{i,\uparrow}^F+ \hat{n}_{i,\downarrow}^F\right)\text{,}
\end{equation}
\noindent where $U_{BF}$ is the boson-fermion interaction parameter.\par 
A numerical study of Hamiltonian~\eqref{eq:HBF} requires fixing the energy scale ($t_F= t_B= 1$), and restricting the number of bosons per site, which is unbounded. It is common to consider the hard-core limit, which implies that each site can be occupied by at most one boson (which implies no interaction term in Eq.~\eqref{eq:HB}), leading to the following local basis: $\frac{|F\rangle}{|B\rangle}=\frac{|0\rangle}{|0\rangle},\frac{|\uparrow\rangle}{|0\rangle},\frac{|\downarrow\rangle}{|0\rangle},\frac{|\uparrow\downarrow\rangle}{|0\rangle},\frac{|0\rangle}{|1\rangle},\frac{|\uparrow\rangle}{|1\rangle},\frac{|\downarrow\rangle}{|1\rangle},\frac{|\uparrow\downarrow\rangle}{|1\rangle}$.\par  
Taking into account the above basis and using DMRG, we were able to explore the ground state of two-color fermions and scalar bosons in a one-dimensional lattice, for which we expected diverse configurations, as shown in Fig.~\ref{fig4}. One possibility is the mixed Mott insulator, where the total number of carriers is commensurate with the lattice (Fig.~\ref{fig4} (a)), while the others drawn are yet to be established. Therefore, the ground-state energy $E(N^{\uparrow}_F,N^{\downarrow}_F,N_{B})$, for $N_B$ bosons, and $N^{\uparrow}_F$ and $N^{\downarrow}_F$ fermions is calculated for different sets of densities and parameters. Knowing how the system reacts to perturbations is crucial, which can be quantified by the bosonic chemical potential $\mu_B=E(Ṇ^{\uparrow}_F,N^{\downarrow}_F,N_B +1)-E(N^{\uparrow}_F,N^{\downarrow}_F,N_B)$. In Fig.~\ref{fig5}, we follow the evolution of the bosonic chemical potential at the thermodynamic limit as the number of bosons increases, considering fermionic density at half-filling ($\rho_F=1$), fermion-fermion repulsion $U_{FF}=3.5$, and two values for the boson-fermion coupling. Note that under the above parameters, there is no possible emergence of the mixed Mott insulator; hence any plateau in the curve implies a new insulator state to be characterized. A monotonous evolution of the bosonic chemical potential can be seen for $U_{BF}=1$ (black dots), which implies that the bosonic excitations are for free. A different scenario happens for $U_{BF}=6$ (red squares), where the continuous growth of the bosonic chemical potential is broken at the bosonic density $\rho_B=1/2$, indicating the emergence of a finite gap to generate bosonic excitations, as can be seen in the inset. The plateau in Fig.~\ref{fig5} establishes a new kind of insulator state for Bose-Fermi mixtures, because the total number of particles is not commensurate with the lattice and the emergence of this insulator depends on interactions. We found that this new noncommensurate insulator fulfills the condition $\rho_B+\tfrac{1}{2}\rho_F=1$, and a possible distribution of carriers was sketched in Fig.~\ref{fig4} (b).\par
\begin{figure}[t] 
\includegraphics[width=18pc]{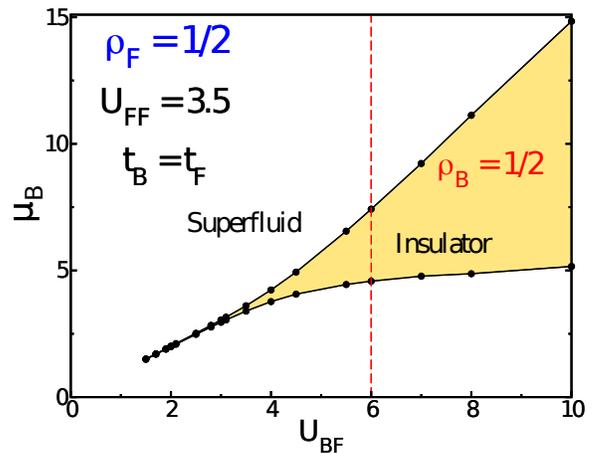} 
\caption{\label{fig6} Bosonic chemical potential versus the boson-fermion repulsion at fermionic half-filling ($\rho_F=1$) and fermion-fermion repulsion $U_{FF}=3.5$. The yellow zone corresponds to an insulator phase for $\rho_B = 1/2$, while the white zone represents a superfluid. The vertical dashed line at $U_{BF}=6$ corresponds to the curve in Fig.~\ref{fig5} (red squares). The lines are visual guides. Adapted from Ref.~\cite{Avella-PRA19}.} 
\end{figure} 
The progress of the noncommensurate insulator as a function of the boson-fermion coupling is depicted in Fig.~\ref{fig6} for a fixed fermion-fermion repulsion $U_{FF}=3.5$ and a fermionic density $\rho_F = 1$. The insulator state (yellow area) is surrounded by the superfluid phase, and there is a critical value from which the insulator emerges and corresponds to $U^{*}_{BF}\approx1.50$. This picture is similar for other values of the fermion-fermion repulsion, although the critical values change.\par 
The interplay between fermions with internal degrees of freedom and scalar bosons led to a new insulator state characterized by the condition $\rho_B+\tfrac{1}{2}\rho_F=1$, but this was for a balanced mixture, i.e. $\rho^{\uparrow}_F=\rho^{\downarrow}_F=\tfrac{1}{2}\rho_F$; therefore, we can surmise that this new insulator involves a commensurability relation between the bosons and one kind of fermion, but what happens to the other kind of fermions? To unvield this scenario, we need to get out of the balanced condition and consider a mixture where $\rho^{\uparrow}_F\neq \rho^{\downarrow}_F$.\par  
\begin{figure}[t]
	\centering
	\includegraphics[width=19pc]{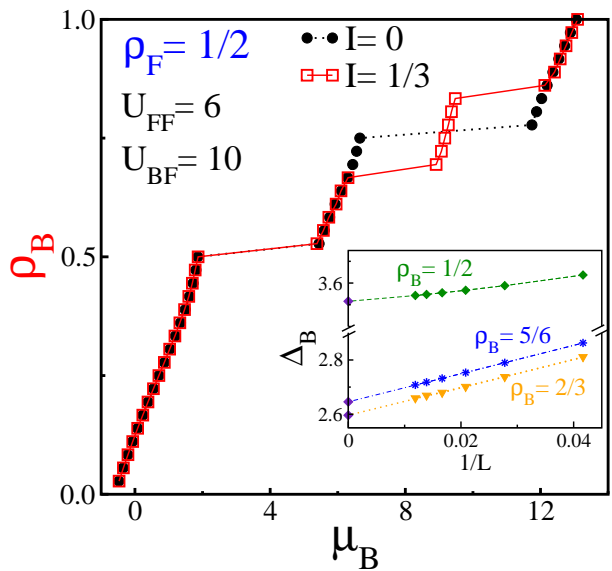}
	\caption{Bosonic density $\rho_B$ versus the bosonic chemical potential $\mu_B$ (thermodynamic limit) for a mixture with a quarter fermionic filling ($\rho_F=1/2$), fermion-fermion repulsion $U_{FF}=6$ and inter-particle interaction $U_{BF}=10$. A balanced $I=0$ (black dots) and an imbalanced mixture $I=1/3$ (red squares) were considered. Inset: Progress of the charge gap associated with the plateau as a function of the inverse of the lattice size, showing that it is nonzero at the thermodynamic limit. The lines are visual guides. Adapted from Ref. ~\cite{GuerreroS-PRA21}.}
	\label{fig7}
\end{figure}
Experimentalists in the cold-atom area can generate asymmetries in the spin populations~\cite{Zwierlein-S06,Partridge-S06,Liao-Nat10,Kinnunen-RPP18, Dobrzyniecki-AQT20}, making such systems ideal for searching for the elusive unconventional pairing mechanism suggested by Fulde, Ferrell, Larkin, and Ovchinnikov (FFLO)~\cite{Fulde64,Larkin65}. In a recent study, it was shown that the visibility of the FFLO state is enhanced as the interparticle strength grows in a Bose-Fermi mixture~\cite{Singh-PRR20}.\par 
In Fig.~\ref{fig7}, we show the thermodynamic limit value for the bosonic chemical as the number of bosons increases from zero for a mixture with a quarter fermionic density ($\rho_F=1/2$), a repulsive fermionic interaction $U_{FF}=6$, and a boson-fermion coupling $U_{BF}= 10$. The black dots correspond to a spin-balanced mixture, for which $\rho^{\uparrow}_F=\rho^{\downarrow}_F=\tfrac{1}{2}\rho_F$, and we observed that the chemical potential grows monotonously for almost all densities, which indicates that there is no cost to generating excitations. However, at the bosonic densities $\rho_B=1/2$ and $\rho_B=3/4$, the chemical potential jumps, generating a plateau in the curve, which is associated with a rise of insulating states. The nature of the insulator state at $\rho_B=1/2$ is obvious, because the total number of carriers will be commensurate with the lattice size, and this state corresponds to the mixed Mott insulator, for which 
$\rho_B+\rho_F=1$, and we emphasize that in a mixture with spinor fermions the mixed Mott insulator also  emerges. The previously-found noncommensurate insulator also arises in this case, at the bosonic density $\rho_B=3/4$, and we corroborate that $\rho_B+\tfrac{1}{2}\rho_F=1$.\par 
\begin{figure}[t]
	\centering
	\includegraphics[width=19pc]{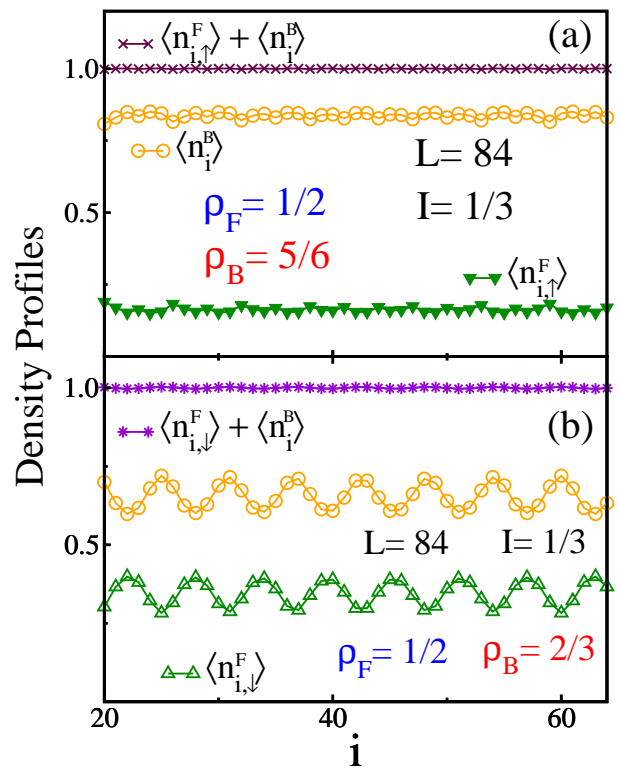}
	\caption{Density profiles of fermions and bosons in a mixture with a quarter fermionic filling ($\rho_F=1/2$) and imbalance parameter $I=1/3$, for which a spin-selective insulator emerges at $\rho_B=5/6$ (a) and $\rho_B=2/3$ (b). The lattice size is $L = 84$, the fermion-fermion repulsion is $U_{FF}=6$, and the inter-species interaction is $U_{BF}=10$. The lines are visual guides. Reproduced from Ref. ~\cite{GuerreroS-PRA21}.}
	\label{fig8}
\end{figure}
To explore a spin-imbalanced mixture, we quantify the difference between the spin populations through $I=(N_F^\downarrow-N_F^\uparrow)/(N_F^\uparrow+N_F^\downarrow)$, and depict the case $I=1/3$ (red squares) in Fig.~\ref{fig7}. We observed that the imbalanced curve matches the balanced one over almost the entire range, indicating that the mixed Mott insulator will emerge regardless of the spin imbalance considered, which is an expected result. However, the main difference between the curves is the disappearance of the insulating state at $\rho_B=3/4$ and the emergence of two new plateaus, which are separated by a superfluid state. The new plateaus are located at the bosonic densities $\rho_B=2/3$ and $\rho_B=5/6$, i.e. one below and one above the noncommensurate insulator of the balanced mixture. We show in the inset of Fig.~\ref{fig7} that there is a finite gap ($\Delta_B=E(N_{F}^{\uparrow},N_{F}^{\downarrow},N_B +1) +E(N_{F}^{\uparrow},N_{F}^{\downarrow},N_B -1)-2E(N_{F}^{\uparrow},N_{F}^{\downarrow},N_B)$) at the thermodynamic for the plateaus, located at $\rho_B=1/2$, $\rho_B=2/3$ and $\rho_B=5/6$, indicating insulator states for these densities. Note that the charge gaps for the plateaus at $\rho_B=2/3$ and $\rho_B=5/6$ are different, which establishes the first difference between these new insulating states.\par
Replicating the imbalanced curve of Fig.~\ref{fig7} for other values of $U_{BF}$, we obtained three insulating lobes surrounded by a superfluid phase, which emerges from different critical values, depending on the fermion-fermion repulsion, the widths of theses lobes increasing further as the boson-fermion coupling grows, similar to what was found in Fig.~\ref{fig6}.\par
\begin{figure}[t] 
\includegraphics[width=18pc]{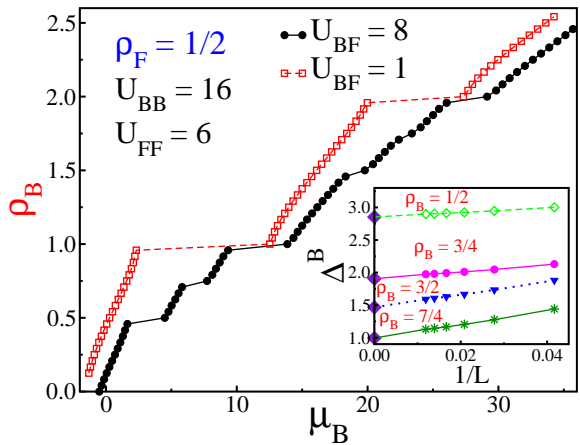} 
\caption{\label{fig9} Progress of the bosonic chemical potential ($\mu_B$) as a function of the bosonic density ($\rho_B$) for a mixture with a fixed fermionic density of $\rho_F = 1/2$ and two values of boson-fermion coupling. Here the boson-boson and fermion-fermion interactions are $U_{BB}=16$ and $U_{FF}=6$, respectively. In the inset, we show that the width of the non-trivial plateaus is finite at the thermodynamic limit. The lines are visual guides.  Adapted from Ref.~\cite{Avella-PRA20}.} 
\end{figure} 
A mixture of two-color fermions and scalar bosons exhibits two insulator states that satisfy $\rho_B+\rho_F=1$ and $\rho_B+\tfrac{1}{2}\rho_F=1$; however, when considering a spin population imbalance, the noncommensurate insulator disappears, giving way to two new insulating states, located at $\rho_B=2/3$ and $\rho_B=5/6$ for $\rho_F=1/2$ and $I=1/3$. The new insulator states are located between the mixed Mott insulator and the trivial one (full lattice), which suggests the relation $\rho_B+\frac{1}{2}\rho_F(1 \pm I)=1$, for locating the new insulators for any fermionic density and imbalanced parameter. Although we found a general relation to locate the new insulators, its physical origin is unclear. To gain more information, we explore the distribution of carriers at these insulators for a finite lattice (Fig.~\ref{fig8}). Considering a lattice with $L=84$ sites, a quarter fermionic density, and a spin population imbalance of $I=1/3$, we display the density profiles of the spin-up fermions and bosons along the lattice for the insulator state with $\rho_B=5/6$ (Fig.~\ref{fig8}(a)). We observe that the expectation value of the local number of both bosons and fermions with spin up slightly oscillates around the values $\left<\hat{n}_i^B \right>\approx 0.833$ and $\left<\hat{n}^{F}_{i,\uparrow} \right>\approx 0.166$, respectively. These curious values lead us to consider the profile of $\left<\hat{n}^{F}_{i,\uparrow} \right>+\left<\hat{n}_i^B \right>$ along the lattice, obtaining one at each site, which suggests a commensurability relation between the bosons and the spin-up fermions. As expected, the local number of fermions with spin down varies around $\left<\hat{n}^{F}_{i,\downarrow} \right>\approx 0.333$ (not shown), and remains in a gapless state, i.e., fixing bosonic density at $\rho_B=5/6$ and $\rho^{\uparrow}_F=1/6$, there is no energy cost to adding spin-down fermions around the density $\rho^{\downarrow}_F=1/3$. In Fig.~\ref{fig8}(b), we show the distribution of the carriers at the insulator state with $\rho_B=2/3$, where a charge density wave is established for each carrier, marking a new difference from the other insulator. We see that the expectation value of spin-down fermions and bosons oscillates off phase and adjusts to meet local commensurability$\left<\hat{n}^{F}_{i,\downarrow} \right>+\left<\hat{n}_i^B \right>=1$, while the spin-up fermions remain in a gapless state.\par 
The above discussion suggests that the new insulator state involves a commensurability relation between the bosons and one kind of fermions, while the other remains in a gapless state. These spin-selective insulators must satisfy the relation $\rho_B+\rho^{\uparrow,(\downarrow)}_F=1$, which is compatible with the relation $\rho_B+\frac{1}{2}\rho_F(1 \pm I)=1$ found previously.\par 
We conclude that a mixture of two-color repulsive fermions and scalar bosons at the hard-core limit exhibits a mixed Mott insulator and two spin-selective insulators that fulfill the commensurability relations $\rho_B+\rho^{\uparrow,(\downarrow)}_F=1$, in a gapless fermion polarized background.\par 
The next question to be resolved is whether spin-selective states emerge if we relax the hard-core condition. It is well-known that the local number of bosons goes from zero to infinity, which forces us to perform a cutoff of the local number of bosons allowed. In our soft-core approximation, we restrict the number of bosons per site to a maximum of $\hat{n}_{max}=3$, which leads us to a large but tractable local Hilbert space of dimension $d=16$, allowing our results to remain unaffected if we increment $\hat{n}_{max}$~\cite{Pai-PRL96,Rossini-NJP12}. Now the interaction term between the bosons will be relevant, and we expect that without the boson-fermion coupling ($U_{BF}=0$), only the well-known Mott insulators (trivial) of each species will appear. In Fig.~\ref{fig9}, we display the progress of the bosonic chemical potential at the thermodynamic limit as the number of bosons increases from zero for a mixture with a quarter fermionic filling ($\rho_F= 1/2$) and repulsion between bosons and fermions of $U_{BB}=16$ and $U_{FF}=6$, respectively. Two values of the boson-fermion interaction $U_{BF}=1$ (open red squares) and $U_{BF}=8$ (closed black dots) were considered, obtaining that $U_{BF}=1$ is a weak coupling between fermions and bosons because the chemical potential evolves continuously, undergoing jumps only at integer bosonic densities, which correspond to bosonic Mott insulators, an expected result in the absence of coupling between fermions and bosons. Comparing Fig.~\ref{fig9} with figures ~\ref{fig5} and ~\ref{fig7} for the hard-core approximation, it is clear that relaxing this condition leads to the emergence of trivial plateaus for larger integer values of the bosonic density. As discussed throughout the paper, increasing the boson-fermion coupling will allow unearthing new discoveries, as can be seen in the closed black dot curve. Clearly, the trivial bosonic Mott insulators survive, but their width is strongly reduced; however, the most important thing is the emergence of two plateaus between trivial Mott insulator ones, which are due to the boson-fermion interaction and are located at the bosonic densities $\rho_B=1/2, 3/4, 3/2,$ and $7/4$. These non-trivial plateaus found for larger boson-fermion repulsion have a finite charge gap ($\Delta_B$) at the thermodynamic limit, as can be seen in the inset, indicating insulator states. Now, we have to classify these states, and we quickly observe that the plateaus located at $\rho_B=1/2$ and $\rho_B=3/2$ correspond to mixed Mott insulators, since they fulfill the relation  $\rho_B+\rho_F=n$, where $n$ is an integer, namely $n=1$ and $2$ for the plateaus at $\rho_B=1/2$ and $3/2$, respectively. Here, we considered a balanced mixture for which $\rho^{\uparrow}_F=\rho^{\downarrow}_F=\tfrac{1}{4}$, and taking into account that the other two non-trivial plateaus are located at $\rho_B=3/4$ and $\rho_B=7/4$, we conclude that the latter insulators are spin-selective ones that satisfy the commensurability relations $\rho_B+\rho^{\uparrow,(\downarrow)}_F=n$ ($n=1,2$), in a gapless fermion polarized background. Therefore, relaxing the hard-core approximation leads us to the emergence of one mixed Mott insulator and one spin-selective one between the trivial Mott insulators in a mixture of two-color fermions and scalar bosons.\par 
It is important to point out that state-of-the-art cold-atom setups allow creating mixtures of two-color fermions and scalar bosons. For instance, mixtures with isotopes $^{171}$Yb and $^{174}$Yb ($^{170}$Yb) have been tested~\cite{YTakasu-JPSJ09}, and dual Bose-Einstein condensates of paired fermions and bosons with $^6$Li and $^7$Li have been achieved experimentally~\cite{Ikemachi-JPB17}. This suggests that the search for the spin-selective insulators reviewed here is an intriguing challenge that can be addressed by experimentalists.\par
\section{\label{sec4}Other possible scenarios}%
Spin-selective insulators emerged initially in the Kondo lattice model, where two kinds of electrons interact, and then were found in mixtures of two-color fermions and scalar bosons; however, we believe that these peculiar insulators can arise in other scenarios, which we suggest below.
\subsection{\label{sec4-1}Periodic Anderson model}%
The standard model for studying the physics of heavy fermion materials is the periodic Anderson model (PAM), whose Hamiltonian is given by
\begin{eqnarray}\label{HPAM} 
\mathcal{\hat{H}}=&-&t\sum_{i,\sigma}\left(\hat{c}^{\dagger}_{i\sigma}\hat{c}_{i+1\sigma}+\hat{c}^{\dagger}_{i+1\sigma}\hat{c}_{i\sigma}\right)+
E_{f}\sum_{i\sigma}\hat{n}^{f}_{i,\sigma} \nonumber 
\\ &+&V\sum_{i,\sigma}\left(\hat{c}^{\dagger}_{i\sigma}\hat{f}_{i\sigma}+\hat{f}^{\dagger}_{i\sigma}\hat{c}_{i\sigma}\right)+
U_{f}\sum_{i}\hat{n}^{f}_{i\uparrow}\hat{n}^{f}_{i\downarrow}, 
\end{eqnarray}
\noindent where $\hat{c}^{\dagger}_{i\sigma}(\hat{f}^{\dagger}_{i\sigma})$ creates an electron in the conduction (localized) band at site $i$ with spin $\sigma= \uparrow, \downarrow$. The local number operator with spin $\sigma$ for the localized electrons is $\hat{n}^{f}_{i\sigma}=\hat{f}^{\dagger}_{i\sigma}\hat{f}_{i\sigma}$. The Coulomb repulsion between two localized electrons at the same site is quantified by $U$, $t$ is the nearest-neighbor hopping integral, and $V$ is the hybridization between the conduction and the localized bands. Finally, $E_f=-U/2$ is the energy displacement of the localized band, which take this value in the symmetric case.\par
\begin{figure}[t] 
\includegraphics[width=18.9pc]{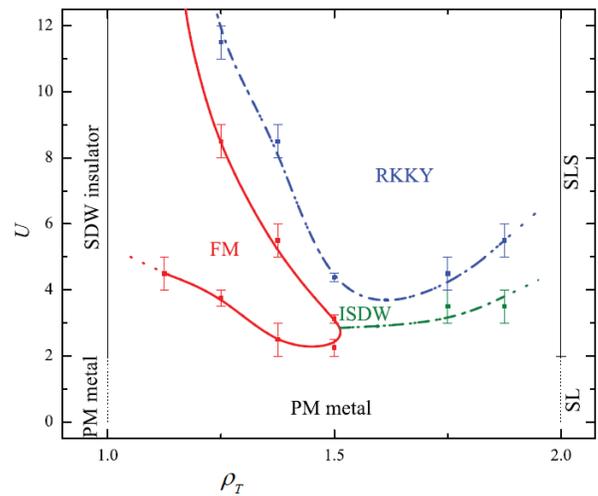} 
\caption{\label{fig10} Phase diagram for the periodic Anderson model considering  electron densities between a quarter and half-filling. The hybridization parameter is fixed at $V/t=0.75$. Taken from ~\cite{Bertussi-PRB11}.} 
\end{figure}

 
Since its appearance, this Hamiltonian~\eqref{HPAM} has been widely studied, using various analytical and numerical techniques; however, we still know little about it: for example, not even in one dimension do we know a complete phase diagram. For the specific case of $V/t=3/4$, Bertussi \textit{et al.} built the phase diagram shown in Fig.~\ref{fig10}, where they considered electronic densities between $\rho_T=\rho_{F,c}+\rho_{F,f}= 1$ (quarter filling) and $\rho_T= 2$ (half filling)~\cite{Bertussi-PRB11}. A metal-insulator transition from a paramagnetic metal to an insulator with an $f$-band spin-density wave was found at quarter filling, whereas at half-filling the ground state is always an insulating spin liquid. Between a quarter and half-filling the ground state is always metallic, exhibiting a very rich magnetic behavior; for instance, we found ferromagnetic, paramagnetic, RKKY, and incommensurate spin density wave (ISDW) regions.\par 
Remembering that the ferromagnetic phase of the Kondo lattice model exhibits peculiar properties, such as that the minority conduction electrons form an insulating state, while the majority conduction electrons form a metallic state, which characterizes the spin-selective insulators, we expect that these insulators also emerge in the ferromagnetic phase of the periodic Anderson model.\par  
\subsection{\label{sec4-2}The ``$g\text{-}e$'' model }%
The possibility of studying the interplay of charge and spin degrees of freedom in clean and fully controllable setups leads to considering the special features of alkaline-earth-like atoms (Yb and Sr) to confine them in traps~\cite{Honerkamp-PRL04,Cazalilla-RPP14,Capponi-AP16}. Because these atoms can be confined in two different optical lattices, the description is given in terms of the two-orbital $SU(N)$-symmetric Hubbard model, more commonly called the ``$g\text{-}e$'' model~\cite{Gorshkov-NP10,Nonne-MPLB11,Bois-PRB15}. This model has allowed extending and revising other models, such as the Kugel-Khomskii model, the Kondo lattice model, and the periodic Anderson model, as well as the discovery of several characteristics of the ``$g\text{-}e$'' model with $N>2$. It has been shown that the high $SU(N)$ symmetry can lead to the emergence of exotic properties in quantum magnetism and pairing superfluidity and to the realization of symmetry-protected topological phase transitions~\cite{Capponi-AP16,Nakagawa-PRB17}.\par 
Alkaline-earth-like atoms have a long-lived metastable excited state $^{3}P_{0}$ ($\arrowvert e\rangle$)  coupled to the ground state $^{1}S_{0}$ ($\arrowvert g\rangle$) via an ultranarrow doubly-forbidden transition, and at low temperatures, the spin-changing collisions are prohibited; hence four different scattering lengths for the states $\arrowvert ee\rangle$,$\arrowvert gg\rangle$, and $\frac{1}{\sqrt{2}}(\arrowvert ge\rangle \pm \arrowvert eg\rangle)$ arise. The description of these atoms confined in optical lattices can be done in terms of the following Hamiltonian ~\cite{Gorshkov-NP10}:
\begin{eqnarray}\label{GEM} 
\mathcal{\hat{H}}=&-&\sum_{m=g,e}t_m\sum_{i,\alpha}\left(\hat{c}^{\dagger}_{m\alpha,i}\hat{c}_{m\alpha,i+1}+ \text{H.c.}\right)\nonumber \\
 &+&\frac{U}{2}\sum_{m=g,e}\sum_i \hat{n}_{m,i}(\hat{n}_{m,i}-1) + V\sum_i \hat{n}_{g,i}\hat{n}_{e,i}\nonumber \\
 &+& V^{g\text{-}e}_{ex}\sum_{i,\alpha,\beta} \hat{c}^{\dagger}_{g\alpha,i} \hat{c}^{\dagger}_{e\beta,i} \hat{c}_{g\beta,i}\hat{c}_{e\alpha,i}, 
\end{eqnarray}
\noindent where, $i$ varies along the sites of a one-dimensional lattice of size $L$, $\hat{c}^{\dagger}_{m\alpha,i}$ creates an atom at site $i$ with orbital index $m=g,e$ and nuclear spin index $\alpha= \uparrow, \downarrow$. The local density operator for each species is $\hat{n}_{m,i}=\sum_{\alpha=\uparrow, \downarrow}\hat{c}^{\dagger}_{m\alpha,i}\hat{c}_{m\alpha,i}$. The strength parameters $U$ and $V$ quantify the local intra- and inter-species interactions, respectively, while $V^{g\text{-}e}_{ex}$ measures the hybridization between the optical lattices. The hopping parameter for each species is $t_m$.\par
Analytical and numerical studies have shown a rich phase diagram at half-filling, where the spin Peierls, charge density wave, orbital density wave, and rung singlet phases predominate in the phase diagram for positive and negative hybridization between the lattices~\cite{Nonne-MPLB11,Bois-PRB15}. We believe that under an adequate set of parameters, the ``$g\text{-}e$'' model will exhibit the spin-selective insulators reviewed here.\par
\subsection{\label{sec4-3}Mixtures of spinor bosons and scalar fermions}%
At the heart of some spectacular phenomena in physics are the  mediated interactions; for instance,  conventional  superconductivity is due to the fact that electrons can form Cooper pairs by interacting via phonons. Also, electrons act as mediators for interactions between magnetic impurities, leading to RKKY  interaction, which is fundamental for the heavy fermion materials. A few years ago, attractive boson–boson interactions mediated by fermions were reported, which are expected to form new magnetic phases and supersolids~\cite{DeSalvo-Nat19,Edri-PRL20}.\par 
Another way to study Bose-Fermi mixtures is to consider spinor bosons and scalar fermions, which have been little studied so far. Recently, it was found that spin-dependent fermion-mediated interactions dramatically modify the properties of binary Bose-Einstein condensates~\cite{RLiao-PRR20}. A mixture of spinor bosons and scalar fermions can be described by the following Hamiltonian:
\begin{eqnarray}\label{HSBSF} 
\mathcal{\hat{H}}=&-&t_B\sum_{i,\sigma}\left(\hat{b}^{\dagger}_{i\sigma}\hat{b}_{i+1\sigma}+\text{H.c.}\right)
+\frac{U^{BB}_0}{2}\sum_{i}\hat{n}_{i}^{B}\left(\hat{n}_{i}^{B}-1\right)\nonumber \\ 
&+&\frac{U^{BB}_2}{2}\sum_{i,\sigma}\left(\hat{b}^{\dagger}_{i\downarrow}\hat{b}^{\dagger}_{i\downarrow}\hat{b}_{i\uparrow}\hat{b}_{i\uparrow}+\hat{b}^{\dagger}_{i\uparrow}\hat{b}^{\dagger}_{i\uparrow}\hat{b}_{i\downarrow}\hat{b}_{i\downarrow}\right)\nonumber \\
&+&(U^{BB}_0+U^{BB}_2)\sum_{i}\hat{n}^{B}_{i\uparrow}\hat{n}^{B}_{i\downarrow}
-t_F \sum_{i}\left(\hat{f}_i^\dagger\hat{f}_{i+1} + \text{H.c.} \right)\nonumber \\
&+&U_{BF}\sum^{L}_{i=1}\hat{n}_i^F\left(\hat{n}_{i,\uparrow}^B+ \hat{n}_{i,\downarrow}^B\right)\text{,}
\end{eqnarray}
\noindent where $\hat{b}_{i,\sigma}$ $(\hat{b}_{i,\sigma}^\dagger)$ annihilates (creates) a boson with spin $\sigma=\uparrow,\downarrow$ at the lattice site $i$, and $\hat{n}^{B}_{i,\sigma}=\hat{b}_{i,\sigma}^{\dag}\hat{b}_{i,\sigma}$ is the local number operator for each kind of bosons, such that $\hat{n}^{B}_{i}=\hat{n}^{B}_{i,\uparrow}+\hat{n}^{B}_{i,\downarrow}$. The operator $\hat{f}_i^\dagger$ ($\hat{f}_i$) creates (annihilates) a fermion at the lattice site $i$, and  $\hat{n}_i^F=\hat{f}_i^\dagger\hat{f}_i$ is the local number operator for fermions. The boson-fermion interaction is quantified for the parameter $U_{BF}$, the hopping of bosons (fermions) is modulated by $t_B$ ($t_F$), and $U^{BB}_0$ quantifies the spin-independent contact repulsion, while $U^{BB}_2$ is the spin-dependent interaction coupling~\cite{Forges-PRB10}.\par 
Exploring the Hamiltonian~\eqref{HSBSF} will allow us to determine the dependence of the spin-selective insulators with the Pauli exclusion principle as well as effective mediated interactions and possible new ground states.\par 

\section{\label{sec5} Conclusions}
This paper presents some scenarios where particular spin-selective insulators arise, which appear in systems composed of two different kinds of carriers, one of them being fermions with two internal degrees of freedom, while the other could be localized spins or scalar bosons. Spin-selective insulators have a finite charge gap, and one kind of fermions remains in an insulator state, while the other one is in a gapless state.\par 
The physical properties of heavy fermion materials involve electrons from two different bands, and the simplest model for describing some of them is the Kondo lattice model, which exhibits a rich phase diagram with Kondo insulator, spiral, ferromagnetic, and island phases. Namely, for the ferromagnetic phase it was found that the cooperation between a partial Kondo screening and ferromagnetism leads to a spin-selective insulator where the majority-spin conduction electrons are in a metallic state while the minority-spin ones remain in an insulator state, fulfilling the commensurability relation $\rho^{\downarrow}_{F,c}+\rho^{\downarrow}_{F,f}=1$.\par
In a balanced mixture of two-color fermions and scalar bosons, two non-magnetic insulators surrounded by superfluids arise. The repulsive character of the interactions causes one of insulators to be a mixed Mott insulator, where the sum of the number of bosons and fermions are commensurate with the lattice, satisfying the relation $\rho_B+\rho_F=n$, where $n$ is an integer. The other insulator corresponds to the spin-selective one, for which the total magnetization is zero and it satisfies the commensurability relation $\rho_B+\rho^{\uparrow}_F=\rho_B+\rho^{\downarrow}_F=n$, in a gapless fermion polarized background. Note that between trivial Mott insulators of the mixture, one mixed Mott insulator and one spin-selective insulator always appear.\par 
If a spin population imbalance is generated in a mixture of two-color fermions and scalar bosons $\rho^{\uparrow}_F\neq \rho^{\downarrow}_F$, the mixed Mott insulator remains unaltered; however, the non-magnetic spin-selective insulator splits into two ferromagnetic spin-selective insulators that fulfill the relations $\rho_B+\rho^{\uparrow}_F=n$ and $\rho_B+\rho^{\downarrow}_F=n$.\par 
Finally, we suggest new scenarios were the spin-selective insulator could arise, such as the periodic Anderson model, the ``$g\text{-}e$'' model, and a mixture of spinor bosons and scalar fermions.\par 
We hope that this contribution may stimulate more investigation of these insulators and the search for them in diverse experimental setups.\par 
\section*{Acknowledgments}
J. Silva-Valencia acknowledges the support od the DIEB- Universidad Nacional de Colombia (Grant No. 51116), and expresses thanks to J. J. Mendoza-Arenas, R. Avella, R. Guerrero-Suarez, and R. Franco.

\bibliography{/home/jereson/PAPERS/Bib/Bibliografia.bib}

\end{document}